\documentclass[twocolumn, showkeys,showpacs, prb, floatfix]{revtex4}
\usepackage{amsfonts, amsmath, amssymb, latexsym}
\usepackage[dvips]{graphicx}
\newtheorem{Pr1}{Proposition}

\begin{document}

\title{Numerical implementation of some reweighted path integral methods}

\author{Cristian Predescu, Dubravko Sabo, and J. D. Doll}

\affiliation{
Department of Chemistry, Brown University, Providence, Rhode Island 02912
}
\date{\today}
\begin{abstract}
The reweighted random series techniques provide finite-dimensional approximations to the quantum density matrix of a physical system that have fast asymptotic convergence. We study two special reweighted techniques that are based upon the L\'evy-Ciesielski and Wiener-Fourier series, respectively.  In agreement with the theoretical predictions, we demonstrate by numerical examples that the asymptotic convergence of the two reweighted methods is cubic for smooth enough potentials. For each reweighted technique, we propose some minimalist quadrature techniques for the computation of the path averages. These quadrature techniques are designed to preserve the asymptotic convergence of the original methods. 
\end{abstract}
\maketitle

\section{Introduction}
Random series representations are among the most general starting points for the derivation of finite approximations to the Feynman-Ka\c{c} formula.\cite{Pre02} The Feynman-Ka\c{c} formula computes the density matrix of a quantum system as an infinite dimensional stochastic integral\cite{Fey65, Sim79} and  constitutes the main ingredient for the computation of quantal thermodynamic properties by Monte Carlo path integral techniques.\cite{Cep95} Current research is focused on the development of rapidly convergent finite-dimensional approximations to the Feynman-Ka\c{c} formula. In deciding upon the efficiency of a certain method, one has to take into account both the number of path variables required  and the computational time necessary to evaluate the various quantities involved. 

One special group of path integral methods, which we shall refer to as direct methods, have the property that they require knowledge of the intermolecular potential only for the computation of the density matrix or the partition function of a physical system.  Numerical algorithms based upon such methods have the advantage that they can be used as ``black boxes.'' Many times,  the potential energy surface of a system is a rather complicated function of the physical coordinates and there are no computational alternatives to such direct methods. 
It is therefore important to look for those direct methods that have the fastest asymptotic convergence and which minimize the number of calls to the potential. 

The most widely used direct method is the trapezoidal Trotter discrete path integral (DPI) method.\cite{Tro59,Rae83} It is usually derived by means of the Trotter product rule and an appropriate short-time high-temperature approximation. The formal asymptotic convergence of the trapezoidal Trotter DPI method and of  related DPI techniques was extensively studied by Suzuki\cite{Suz91, Suz85} and was found to be $O(1/n^2)$.\cite{Footnote} As shown in Ref.~(\onlinecite{Pre02b}), such discrete path integral techniques can also be interpreted as direct discretizations of the Feynman-Ka\c{c} formula by means of certain quadrature rules. As a consequence, the dimensionality of the integral approximations for the density matrix is equal to the number of quadrature points. 

A second approach to constructing finite-dimensional approximations for density matrices is based on the random series representations of the Feynman-Ka\c{c} formula. Special modifications of these random series by means of the so-called reweighted techniques have been  initially considered on an intuitive basis in Refs.~(\onlinecite{Pre02}) and (\onlinecite{Pre02b}).  Depending upon the specific series, these modifications have been shown numerically to have asymptotic convergence equal to or faster than that of the trapezoidal Trotter DPI method. More recently, the correct definition and the convergence properties of the reweighted techniques have been shown to depend upon the asymptotic convergence of the partial averaging method.\cite{Pre03} In fact, once the convergence of the partial averaging method has been fully understood,\cite{Pre03a} it became possible to develop reweighted techniques that have optimal asymptotic convergence.\cite{Pre03} The numerical implementation of two optimal reweighed methods that are based upon the L\'evy-Ciesielski and the Wiener-Fourier series, respectively will be thoroughly analyzed in the present article. 

Though their mathematical justification is rather involved, these two reweighted techniques have the advantages that they are direct methods and that they have cubic convergence for smooth enough potentials. However, as with the full Feynman-Ka\c{c} formula, the path averages of the potential,  which involve one-dimensional integrals on a compact interval, are assumed to be computed exactly. In practice, one is forced to make recourse to a certain quadrature scheme. Let us remember that for the trapezoidal Trotter DPI method the number of quadrature points is equal to the number of path variables. It follows that in order to preserve the overall cubic convergence of the reweighted methods, one needs to utilize a quadrature technique which uses a number of quadrature points proportional to the number of path variables and which preserves the asymptotic convergence of the reweighted techniques. It is within the scope of the present article to show that such quadrature schemes exist and to construct minimalist examples for either of the two reweighted techniques presented. 

The main tasks of the present article are a) to review in a transparent fashion the construction of the reweighted techniques, b) to present numerical examples demonstrating the cubic convergence of the methods, and c) to devise quadrature schemes that preserve  the asymptotic convergence of the methods, while employing a number of quadrature points proportional to the number of path variables. Based upon the numerical evidence presented as well as on the formal convergence properties of the methods, we eventually conclude that the reweighted techniques are the most efficient direct methods currently available.  As such, these techniques constitute the methods of choice for use in path integral simulations.

\newcommand{\ud}{\mathrm{d}}
\section{Reweighted random series methods}
In this section, we give a short review of the path integral methods we study in this paper. According to Ref.~(\onlinecite{Pre02}), the most general random series representation for the Feynman-Ka\c{c} formula can be constructed as follows. Assume given $\{\lambda_k(\tau)\}_{k \geq 1}$ a system of functions on the interval $[0,1]$, which together
with the constant function $\lambda_0(\tau)=1$ makes up an orthonormal basis in $L^2[0,1]$.
  Let 
\[ \Lambda_k(t)=\int_0^t \lambda_k(u)\ud u.\]
If $\Omega$
is the space of infinite sequences $\bar{a}\equiv(a_1,a_2,\ldots)$ and
\begin{equation}
\label{eq:1}
\ud P[\bar{a}]=\prod_{k=1}^{\infty}\ud \mu(a_k)
\end{equation}
  is the probability measure on $\Omega$ such that the coordinate maps 
$\bar{a}\rightarrow a_k$ are independent identically distributed 
(i.i.d.) variables with distribution probability
\begin{equation}
\label{eq:2}
\ud \mu(a_i)= \frac{1}{\sqrt{2\pi}} e^{-a_i^2/2}\,\ud a_i,
\end{equation}
then the Feynman-Ka\c{c} formula has the form
\begin{eqnarray}
\label{eq:3}
  \frac{\rho(x, x' ;\beta)}{\rho_{fp}(x, x' 
;\beta)}&=&\int_{\Omega}\ud P[\bar{a}]\nonumber  \exp\bigg\{-\beta 
\int_{0}^{1}\! \!  V\Big[x_r(u) \\& +& \sigma \sum_{k=1}^{\infty}a_k 
\Lambda_k(u) \Big]\ud u\bigg\}.
\end{eqnarray}
In the above equation, $x_r(u)=x+(x'-x)u$ is the  reference path, $\sigma= (\hbar^2\beta  /m_0)^{1/2}$, and $\rho_{fp}(x,x';\beta)$ denotes the density matrix for a similar free
particle. For a multidimensional system, the Feynman-Ka\c{c} formula is obtained by employing an independent random series for each additional degree of freedom. 

The primitive methods of order $n$ are obtained by truncating the series appearing in the above formula to the rank $n$: 
\begin{eqnarray}
\label{eq:4}
  &&\frac{\rho_n^{\text{Pr}}(x, x' ;\beta)}{\rho_{fp}(x, x' 
;\beta)}=\int_{\mathbb{R}}\ud \mu(a_1)\ldots \int_{\mathbb{R}}\ud 
\mu(a_n)\nonumber \\ &&\times \exp\bigg\{-\beta \; 
\int_{0}^{1}\! \!
V\Big[x_r(u)+\sigma \sum_{k=1}^{n}a_k
\Lambda_k(u) \Big]\ud u\bigg\}. \qquad
\end{eqnarray}
It is known that the primitive methods can achieve at most $O(1/n)$ asymptotic convergence\cite{Pre02} and therefore they are of limited use. This is true especially because there are alternative methods capable of attaining faster asymptotic convergence without any additional work. Such methods include the so-called reweighted random series techniques, which will be discussed in the remainder of the present section. 

According to  Ref.~(\onlinecite{Pre03}), a reweighted method constructed from the random series $\sum_{k=1}^\infty a_k \Lambda_k(u)$ is any sequence of approximations to the density matrix of the form
\begin{eqnarray}
\label{eq:5}&&
\frac{\rho^{\text{RW}}_n(x, x' ;\beta)}{\rho_{fp}(x, x' 
;\beta)}=\int_{\mathbb{R}}\ud \mu(a_1)\ldots \int_{\mathbb{R}}\ud 
\mu(a_{qn+p})\nonumber  \\&& \times \exp\bigg\{-\beta \; \int_{0}^{1}\! \!
V\Big[x_r(u)+ \sigma \sum_{k=1}^{qn+p}a_k \tilde{\Lambda}_{n,k}(u)\Big]\ud u\bigg\},\qquad
\end{eqnarray}
where $q$ and $p$ are some fixed integers, where
\begin{equation}
\label{eq:6}
\tilde{\Lambda}_{n,k}(u)= \Lambda_k(u) \quad \text{if} \ 1\leq k \leq n, 
\end{equation} 
and where
\begin{equation}
\label{eq:7}
\sum_{k=n+1}^{qn+p}\tilde{\Lambda}_{n,k}(u)^2=\sum_{k=n+1}^{\infty}\Lambda_{k}(u)^2.
\end{equation}
Notice that the expression given by Eq.~(\ref{eq:5}) is quite general and comprises the primitive methods by setting $\tilde{\Lambda}_{n,k}(u)=0$ for $k \geq n+1$. Of course, in this case Eq.~(\ref{eq:7}) no longer holds. 

The reweighed techniques are not uniquely defined by the  series representations on which they are based. There  are an infinite number of reweighted techniques associated with a given series. Some are better, others are worse. However, they share the common property that their asymptotic convergence is known.\cite{Pre03} The respective convergence theorems  constitute definite criteria for the  selection of optimal reweighted techniques. Two such optimal methods will be presented in the remainder of the present section. 

\subsection{A Wiener-Fourier reweighted technique}
The Wiener-Fourier series representation is based on the system of functions $ \{\lambda_k(\tau) =\sqrt{2} \cos(k 
\pi \tau);\; k \geq 1\}$,  which together with the constant function makes up a complete orthonormal system. In this case, we have 
\[
\Lambda_k(t)= \int_0^t \sqrt{2} \cos(k \pi \tau) \ud \tau = \sqrt{\frac{2}{\pi^2}} \frac{\sin(k\pi t)}{k}.
\]
It has been shown\cite{Pre02} that the Wiener-Fourier series representation is the optimal series representation as to the minimization of the number of variables used to parameterize the paths. This has been argued to be true of the primitive and partial averaging techniques, but it is also true of the reweighted techniques. 

A reweighted method based upon the Wiener-Fourier series representation and having superior asymptotic convergence can be constructed as follows.\cite{Pre03} We let $q=4$ and $p=0$ in Eq.~(\ref{eq:5}) and define
\begin{equation}
\label{eq:8}
r_n(u)=\sqrt{\frac{\gamma_n(u,u)}{{\gamma}^\circ_n(u,u)}}, 
\end{equation}
where
\begin{eqnarray}
\label{eq:9}&&
\gamma_n(u,u)=\frac{2}{\pi^2}\sum_{k=n+1}^{\infty}\frac{ \sin(k\pi u)^2}{k^2}\nonumber \\&&=
u(1-u) - \frac{2}{\pi^2}\sum_{k=1}^{n}\frac{ \sin(k\pi u)^2}{k^2},\quad
\end{eqnarray}
\begin{equation}
\label{eq:10}
{\gamma}_n^\circ(u,u)=\frac{2\alpha_n}{3n}\sum_{k=n+1}^{4n} \sin(k\pi u)^2,
\end{equation}
and
\begin{equation}
\label{eq:11}
\alpha_n= \int_0^1 \gamma_n(u,u)\ud u=\frac{1}{\pi^2}\sum_{k=n+1}^\infty \frac{1}{k^2}=\frac{1}{6}-\frac{1}{\pi^2}\sum_{k=1}^n \frac{1}{k^2}. 
\end{equation}
Next, we let
\begin{equation}
\label{eq:12}
\tilde{\Lambda}_{n,k}(u)=
\sqrt{\frac{2}{\pi^2}}  \frac{\sin(k \pi u)}{k}
\end{equation}
for $1 \leq k \leq n$ and
\begin{equation}
\label{eq:13}
\tilde{\Lambda}_{n,k}(u)=
\sqrt{\frac{2\alpha_n}{3n}}r_n(u)\sin(k\pi u)
\end{equation}
for  $n<k \leq 4n$.
One easily verifies that
\begin{eqnarray*}
 \sum_{k=n+1}^{4n}\tilde{\Lambda}_{n,k}(u)^2 = \frac{2\alpha_n}{3n}r_n(u)^2  \sum_{k=n+1}^{4n}\sin(k\pi u)^2\\  = {\gamma}_n(u,u)= \sum_{k=n+1}^{\infty}\Lambda_k(u)^2.
\end{eqnarray*}

Clearly, the method defined by Eqs.~(\ref{eq:5}), (\ref{eq:12}), and (\ref{eq:13}) is  a reweighted technique derived from the Wiener-Fourier series representation. Its explicit formula is 
\begin{eqnarray}
\label{eq:14}
  &&\frac{\rho_n^{\text{WF}}(x, x' ;\beta)}{\rho_{fp}(x, x' 
;\beta)}=\int_{\mathbb{R}}\ud \mu(a_1)\ldots \int_{\mathbb{R}}\ud 
\mu(a_{4n})\nonumber \\ \nonumber &&\times \exp\Bigg\{-\beta \; 
\int_{0}^{1}\! \!
V\Bigg[x_r(u)+\sigma \sum_{k=1}^{n}a_k
\sqrt{\frac{2}{\pi^2}} \frac{\sin(k\pi u)}{k}\\ && + \sigma \sqrt{\frac{2\alpha_n}{3n}} r_n(u) \sum_{k=n+1}^{4n}a_k
\sin(k\pi u) \Bigg]\ud u\Bigg\}.
\end{eqnarray}

The reweighted technique defined above will be denoted by RW-WFPI even if, strictly speaking, this acronym should denote the class of all Wiener-Fourier reweighted methods. Even if it might not be the best reweighted technique in its class, this particular RW-WFPI technique has good asymptotic convergence, as described by the following convergence results. Assume that $V(x)$ has finite Gaussian transform and  has \emph{second} order Sobolev derivatives [for the exact conditions in which the results hold, the reader should consult Ref.~(\onlinecite{Pre03})]. Then,
\begin{eqnarray}
\label{eq:15}\nonumber 
\lim_{n \to \infty} n^3 
\left[\rho(x,x';\beta)-\rho_n^{\text{WF}}(x,x';\beta)\right]
 = \frac{\hbar^2\beta^3}{N_{0}\pi^4 m_0}\\ \times  
\rho(x,x';\beta)\left[V'(x)^2+V'(x')^2\right], \qquad
\end{eqnarray}
where 
\begin{equation}
\label{eq:16}
N_{0}=15.0045\ldots
\end{equation}
In the case of a $d$-dimensional system, the convergence constant becomes
\begin{eqnarray}
\label{eq:17}\nonumber 
\lim_{n \to \infty} n^3 
\left[\rho(x,x';\beta)-\rho_n^{\text{WF}}(x,x';\beta)\right]
 = \frac{\hbar^2\beta^3}{N_{0}\pi^4}\\ \times  
\rho(x,x';\beta)\left\{\sum_{i=1}^d \frac{[\partial_i V(x)]^2+[\partial_i V(x')]^2}{m_{0,i}}\right\}, 
\end{eqnarray}
where $\partial_i V(x)$ denotes the first order partial derivative $\partial V(x)/ \partial x_i$.

\subsection{ A L\'evy-Ciesielski reweighted method}

\begin{figure}[!tbp] 
   \includegraphics[angle=270,width=8.5cm,clip=t]{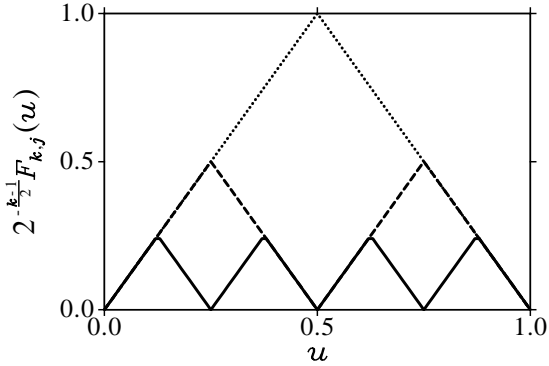} 
 \caption[sqr]
{\label{Fig:1}
A plot of the renormalized Schauder functions for the layers $k=1,2,\,\text{and}\,3$ showing the pyramidal structure.
}
\end{figure}

Another important representation of the Feynman-Ka\c{c} formula is based upon the so-called L\'evy-Ciesielski series,\cite{Pre02b} which is constructed as follows. For $k=1,2,\ldots$ and $j=1,2,\ldots,2^{k-1}$, the Haar function $f_{k,j}$  is defined by
\begin{equation}
\label{eq:18}
f_{k,j}(t)=\left\{\begin{array}{cc} 2^{(k-1)/2},& t \in [(l-1)/2^k, l/2^k]\\ - 2^{(k-1)/2},& t \in [l/2^k, (l+1)/2^k]\\ 0, &\text{elsewhere,} \end{array}\right.
\end{equation}
where $l=2j-1$.
Together with $f_0\equiv 1$, these functions make up a complete orthonormal basis in $L^2([0,1])$. Their primitives 
\begin{widetext}
\begin{equation}
\label{eq:19}
F_{k,j}(t)=\left\{\begin{array}{cc} 2^{(k-1)/2}[t-(l-1)/2^k],& t \in [(l-1)/2^k, l/2^k]\\ 2^{(k-1)/2}[(l+1)/2^k-t],& t \in [l/2^k, (l+1)/2^k]\\ 0, &\text{elsewhere} \end{array}\right.
\end{equation}
\end{widetext}
are called the \emph{Schauder functions}. The Schauder functions resemble some ``little tents'' that can be obtained one from the other by dilatations and translations. 
More precisely, we have
\begin{equation}
\label{eq:20}
F_{k,1}(u)=2^{-(k-1)/2}F_{1,1}(2^{k-1}u)
\end{equation}
for $k \geq 1$ and 
\begin{equation}
\label{eq:21}
F_{k,j}(u)=F_{k,1}\left(u-\frac{j-1}{2^{k-1}}\right)
\end{equation}
for $k\geq 1$ and $ 1\leq j \leq 2^{k-1}$. The scaling relations above have to do with the fact that the original Haar wavelet basis makes up a multiresolution analysis of $L^2([0,1])$  organized in ``layers'' indexed by $k$. If we disregard the factor $2^{(k-1)/2}$, the Schauder functions make up a pyramidal structure as shown in Fig.~\ref{Fig:1}.
Let us notice that any reasonable  L\'evy-Ciesielski primitive method should use a total of $n=2^k -1$ Schauder functions, corresponding to $k$ complete layers. We shall assume that $n=2^k -1$ for the remainder of the section. 

According to Ref.~(\onlinecite{Pre03}), a L\'evy-Ciesielski reweighted method having superior asymptotic convergence can be constructed with the help of $3\cdot 2^k= 3n+3$ additional functions organized in two layers of $2^k$ and $2^{k+1}$ functions, respectively. These additional functions are  constructed as follows. 
First, we define a function $r(u)$ which is zero outside the interval $(0,1)$ and equal to
\begin{equation}
\label{eq:22}
\left\{\frac{u(1-u)}{w^2F_{1,1}(u)^2+\left(4-2w^2\right)\left[F_{2,1}(u)^2+F_{2,2}(u)^2\right]}\right\}^{1/2}
\end{equation}
on the interval $(0,1)$. The constant $w$ appearing in Eq.~(\ref{eq:22}) has the value
\begin{equation}
\label{eq:23}
w=0.62258\ldots
\end{equation}
Next, we define the functions: 
\[C_{0}(u)= w r(u) F_{1,1}(u),\]
\[L_{0}(u)= \sqrt{4-2w^2} r(u) F_{2,1}(u),\]
and
\[R_{0}(u)= \sqrt{4-2w^2} r(u) F_{2,2}(u).\]

\begin{figure}[!tbp]
   \includegraphics[angle=270,width=8.5cm,clip=t]{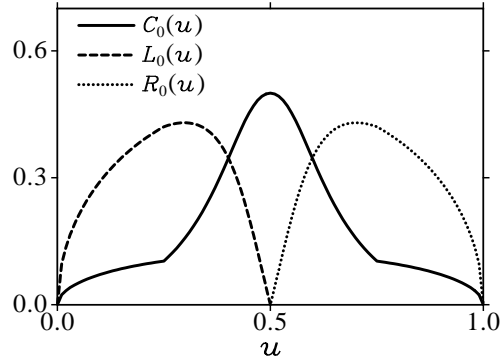}
  \caption[sqr]
{\label{Fig:2}
Shapes of the functions $C_0(u)$, $L_0(u)$, and $R_0(u)$ utilized for the the construction of the reweighted technique. }
\end{figure}

The functions  $C_0(u)$, $L_0(u)$, and $ R_0(u)$ are plotted in Fig.~\ref{Fig:2}. 
They vanish outside the intervals $[0,1]$, $[0,1/2]$, and $[1/2,1]$, respectively. From them, by dilatations and translations, we construct 
\begin{eqnarray}
\label{eq:24} \nonumber
C_{k,1}(u)= 2^{-k/2}C_{0}\left( 2^k u \right), \quad L_{k,1}(u)=2^{-k/2} L_{0}\left(2^k u \right), \\ R_{k,1}(u)=2^{-k/2} R_{0}\left(2^k u \right), \qquad \qquad \qquad
\end{eqnarray}
as well as
\begin{eqnarray}
\label{eq:25} \nonumber
C_{k,j}(u)= C_{k,1}\left[u-(j-1)/2^{k}\right],\\ L_{k,j}(u)= L_{k,1}\left[u-(j-1)/2^{k}\right], \\ R_{k,j}(u)= R_{k,1}\left[u-(j-1)/2^{k}\right], \nonumber  
\end{eqnarray}
for $1\leq j \leq 2^k$.

In order to have a more compact notation, we set (remember that $n = 2^k -1$)
\[
\tilde{F}^{(n)}_{l, j}(u)= F_{l,j}(u)
\]
for $ 1\leq l \leq k$ and  $\ 1\leq j \leq 2^{l-1}$,
\[
\tilde{F}^{(n)}_{k+1, j}(u)= C_{k,j}(u) 
\]
for $ \ 1\leq j\leq 2^k$, and
\[
\tilde{F}^{(n)}_{k+2, j}(u)= \left\{ \begin{array}{l l} 
L_{k,p}(u),&  \text{if}\ j=2p-1,\\
 R_{k, p}(u), & \text{if}\ j=2p,\end{array} \right.
\]
for $1 \leq j \leq 2^{k+1}$.
Then, the reweighted technique has the explicit form
\begin{widetext}
\begin{eqnarray}
\label{eq:26} \nonumber
\frac{\rho_n^{\text{LC}}(x, x' ;\beta)}{\rho_{fp}(x, x' 
;\beta)}=&&\int_{\mathbb{R}}\ud a_{1,1}\ldots \int_{\mathbb{R}}\ud a_{k+2,2^{k+1}}  \left( 2\pi \right)^{-(4n+3)/2}  \exp\left({-\frac{1}{2}\sum_{l=1}^{k+2}\sum_{j=1}^{2^{l-1}}  a_{l,j}^2}\right)
\\&& \times \exp\left\{-\beta \int_0^1 V\left[x_r(u)+\sigma \sum_{l=1}^{k+2} a_{l,[2^{l-1} u]+1} \;\tilde{F}^{(n)}_{l,[2^{l-1} u]+1}(u)\right]\ud u\right\}, 
\end{eqnarray}
\end{widetext}
where $[2^{l-1} u]$ is the integer part of $2^{l-1} u$.

The particular reweighted L\'evy-Ciesielski method described in the present subsection will be denoted by RW-LCPI despite the fact that this acronym should refer to the whole class of reweighted techniques based upon the L\'evy-Ciesielski series. The method was proven to have $o(1/n^2)$ convergence for potentials having first order Sobolev derivatives. It was conjectured that the asymptotic rate of convergence may reach $O(1/n^3)$ for potentials having second order Sobolev derivatives. As a numerical advantage, we notice that the computation of the path at a point $u$ involves  only $k+2=2+\log_2(n+1)$ multiplications and  $k+1=1+\log_2(n+1)$ additions.

\section{Discrete path integral methods}
Usually, by a discrete path integral method one understands an approximation to the density matrix having the Trotter product form
\begin{eqnarray}
\label{eq:27}
\rho_n^{\text{DPI}}(x,x';\beta)=\int_{\mathbb{R}}\ud x_1 \ldots \int_{\mathbb{R}}\ud x_n\; \rho_0\left(x,x_1;\frac{\beta}{n+1}\right)\nonumber \\ \ldots \rho_0\left(x_n,x';\frac{\beta}{n+1}\right),\qquad
\end{eqnarray}
where $ \rho_0\left(x,x';\beta\right)$ stands for a so-called short-time high-temperature approximation. 

In this paper, we shall employ for comparison purposes the trapezoidal Trotter-Suzuki short-time approximation given by the formula
\begin{equation}
\label{eq:28}
\rho_0^{\text{TT}}(x,x';\beta) = \rho_{fp}(x, x';\beta) \exp\left[-\beta \frac{V(x)+V(x')}{2} \right].
\end{equation}
The resulting method will be denoted by the acronym TT-DPI. It has been shown\cite{Pre02b} that for $n=2^k-1$, the TT-DPI method admits the following  fast implementation
\begin{widetext}
\begin{eqnarray}
\label{eq:29}
\frac{\rho_n^{\text{TT}}(x, x' ;\beta)}{\rho_{fp}(x, x' 
;\beta)}=\int_{\mathbb{R}}\ud a_{1,1}\ldots \int_{\mathbb{R}}\ud a_{k,2^{k-1}}  \left( 2\pi \right)^{-n/2}  \exp\left({-\frac{1}{2}\sum_{l=1}^k\sum_{i=1}^{2^{l-1}}  a_{l,i}^2}\right) \\ \times \exp\left\{-{\beta}\sum_{i=0}^{2^k}\omega_i V\left[x_r(u_i)+\sigma \sum_{l=1}^{k}F_{l,[2^{l-1} u_i]+1}(u_i)a_{l,[2^{l-1} u_i]+1}\right]\right\}, \nonumber
\end{eqnarray}
\end{widetext}
where $u_i= 2^{-k} i $ for $0\leq i \leq 2^k$ and 
\[
\omega_i =\left\{ \begin{array}{l l}2^{-(k+1)},& \text{if}\ i\in \{0, 2^k\},\\
2^{-k}, & \text{if}\ 1\leq i \leq 2^k-1.
 \end{array} \right.
\]
The advantage of this scheme is the $\log(n+1)$ scaling of the number of operations one needs to perform in order to compute the path at a discretization point $u_i$. This is accomplished without the use of a fast Fourier transform routine. The asymptotic convergence of the method is known to be $O(1/n^2)$ for smooth enough potentials.\cite{Rae83} 

A second discrete path integral method we consider in this paper is related to the RW-LCPI method. It has been shown that for $n=2^k-1$, the RW-LCPI method can be put in the Trotter product form\cite{Pre03}
\begin{eqnarray}
\label{eq:30}
\rho_n^{\text{LC}}(x,x';\beta)=\int_{\mathbb{R}}\ud x_1 \ldots \int_{\mathbb{R}}\ud x_n\; \rho_0^{\text{LC}}\left(x,x_1;\frac{\beta}{n+1}\right)\nonumber \\ \ldots \rho_0^{\text{LC}}\left(x_n,x';\frac{\beta}{n+1}\right),\qquad
\end{eqnarray}
where
\begin{eqnarray}\nonumber
\label{eq:31}
\frac{\rho_0^{\text{LC}}(x,x';\beta)}{\rho_{fp}(x,x';\beta)}=\frac{1}{\left(2\pi\right)^{3/2}}\int_{\mathbb{R}}\int_{\mathbb{R}}\int_{\mathbb{R}} e^{-\left(a_1^2+a_2^2+a_3^2\right)/2}\\\times \exp\bigg\{-\beta \int_0^1 V[x+(x'-x)u+a_1\sigma C_0(u)\\ +a_2 \sigma L_0(u)+ a_3 \sigma R_0(u)] \ud u\bigg\} \ud a_1 \ud a_2 \ud a_3. \nonumber
\end{eqnarray}
However, for arbitrary $n$, the right-hand side of Eq.~(\ref{eq:30}) defines a new quantity which we shall denote by $\rho^{\text{LC}}_n(x,x';\beta)$, too. Because the $n=2^k-1$ subsequence has $o(1/n^2)$ convergence, it is not  farfetched to assume that the entire $\rho^{\text{LC}}_n(x,x';\beta)$ sequence has $o(1/n^2)$ convergence. We shall verify this assumption later with the help of some numerical examples. 

One of the main advantages of the discrete path integral methods consists of the fact that for low dimensional systems the evaluation of the density matrix and related properties can be performed accurately by means of the numerical matrix multiplication (NMM) method.\cite{Kle73, Thi83} We shall use the NMM method to compute  $n$-th order approximations to the partition function of the type
\[
Z_n^{\text{LC}}(\beta) = \int_{\mathbb{R}}\rho_n^{\text{LC}}(x,x;\beta) \ud x, 
\]
for one-dimensional systems. The main steps of the algorithm are as follows. First, one restricts the system to an interval $[a, b]$ and considers a division of the interval of the type 
\[
x_i = a + i(b-a)/M, \quad 0 \leq i \leq M. 
\]
Next, one computes and stores the symmetric square  matrix of entries
\[A_{i,j} = \frac{b-a}{M} \rho_0^{\text{LC}}\left(x_i, x_j ;\frac{\beta}{n+1}\right), \quad 0\leq i, j \leq M.\] The value of the partition function can then be recovered as 
\[
Z_n^{\text{LC}}(\beta) = \text{tr}\left(A^{n+1}\right).
\]
By computer experimentation, the interval $[a, b]$ and the size $M$ of the division are chosen such that the computation of the partition function is performed with the required accuracy. A fast computation of the powers of the matrix $A$ can be achieved by exploiting the rule $A^{m+n}=(A^m)^n$.  For more details, the reader is referred to the cited literature.  

The use of the NMM technique together with the trapezoidal Trotter approximation raises no special computational problems. However, for the DPI method derived from the L\'evy-Ciesielski reweighted technique, there is the problem of evaluating the short-time approximation given by Eq.~(\ref{eq:31}). Noticing that the function $L_0(u)$ vanishes outside the interval $[0,1/2]$ while the function $R_0(u)$ vanishes outside the interval $[1/2, 1]$, one may recast Eq.~(\ref{eq:31}) in the form
\begin{widetext} 
\begin{eqnarray}\nonumber
\label{eq:32}
\frac{\rho_0^{\text{LC}}(x,x';\beta)}{\rho_{fp}(x,x';\beta)}=\int_{\mathbb{R}} \ud \mu(a_1)\Bigg[ \int_{\mathbb{R}}\ud \mu(a_2)  \exp\bigg\{-\beta  \int_0^{1/2} V[x +(x'-x)u+a_1\sigma C_0(u)  +a_2 \sigma L_0(u)] \ud u\bigg\} \Bigg]\times \\ \Bigg[ \int_{\mathbb{R}}\ud \mu(a_2)  \exp\bigg\{-\beta \int_{1/2}^{1} V[x+(x'-x)u+a_1\sigma C_0(u)  +a_2 \sigma R_0(u)] \ud u\bigg\} \Bigg],  
\end{eqnarray}
\end{widetext}
which is computationally less expensive because it no longer involves a three dimensional Gaussian integral. The Gaussian integrals appearing in Eq.~(\ref{eq:32}) [remember Eq.~(\ref{eq:2})] can be evaluated by means of the Gauss-Hermite quadrature technique.\cite{Pre92} For the purpose of establishing the asymptotic convergence of the partition functions, we have found that a number of $10$ quadrature points for each dimension is sufficient. This is so because the errors due to the Gauss-Hermite quadrature approximation quickly vanish as $\beta / (n + 1) \to 0$.

\section{Path integral quadrature techniques}
The basic assumption utilized in the construction of the reweighted techniques is that the path averages of the type 
\[
\int_{0}^{1}\! \!
V\Big[x_r(u)+ \sigma \sum_{k=1}^{qn+p}a_k \tilde{\Lambda}_{n,k}(u)\Big]\ud u
\]
are computed exactly. In practice, however, one has to perform the integrals approximately by quadrature techniques.  In this section, we shall give a positive answer to the question: Are there any quadrature rules that preserve the fast asymptotic convergence of the reweighted methods while employing at most $q'n+p'$ quadrature points for some fixed integers $q'$ and $p'$? 

It is not within the scope of the present paper to solve this problem in a mathematically rigorous way. Rather, we shall use numerical examples to demonstrate our findings. The key observation toward the solution of the quadrature problem is the fact that the asymptotic rates of convergence of the reweighted and partial averaging techniques depend solely on the first and the second order derivatives of the potential, through their square. Examples are furnished by the convergence constants of the partial averaging Wiener-Fourier method (Theorem~4 of Ref.~\onlinecite{Pre03a}), the partial averaging L\'evy-Ciesielski method (Theorem~4 Ref.~\onlinecite{Pre02b}), or the Wiener-Fourier reweighted method discussed in Section~II.A.

This observation suggests that any quadrature rule capable of integrating exactly or to a good approximation any quadratic potential will preserve the cubic convergence of the respective methods for all smooth enough potentials (potentials having continuous second order derivatives). In the present section we shall focus our attention on finding the fastest quadrature techniques that integrate the quadratic potential. The fact that the cubic convergence of the reweighted methods extends to all sufficiently smooth potentials is then demonstrated in the following sections with the help of some numerical examples. 

\subsection{Minimalist quadrature rule for the RW-LCPI method}

In this subsection, we shall find an explicit quadrature technique that integrates exactly all the quadratic potentials for the RW-LCPI method and its discrete path integral version. We look for a quadrature technique that utilizes a minimal number of quadrature points and which also preserves the intrinsic symmetries of the system of functions utilized for path expansion. 

We begin by noticing that on each interval $\left[(j-1)2^{-k}, j2^{-k}\right]$ the paths appearing in the RW-LCPI formulation  are linear combinations of the functions $1$, $u$, $C_{n,j}(u)$, $L_{n,j}(u)$, and $R_{n,j}(u)$.
That is, they are of the form 
\[
c_1 + c_2 u + c_3 C_{n,j}(u) + c_4 L_{n,j}(u) + c_5 R_{n,j}(u)
\]
with some appropriate values for the constants $c_1, \ldots, c_5$. This is so because the functions $F_{l,p}$ are linear on the intervals $\left[(j-1)2^{-k}, j2^{-k}\right]$ for all $1\leq l\leq k$ and $1\leq p \leq 2^{l-1}$. Using this observation and the  decomposition
\begin{eqnarray*}
\int_0^1 V\Bigg[x_r(u)+\sigma \sum_{l=1}^{k+2} a_{l,[2^{l-1} u]+1} \;\tilde{F}^{(n)}_{l,[2^{l-1} u]+1}(u)\Bigg]\ud u \\ = \sum_{i = 1}^{2^k}\int_{(i-1)2^{-k}}^{i2^{-k}}  V\Bigg[x_r(u) +\sigma \sum_{l=1}^{k+2} a_{l,[2^{l-1} u]+1} \\ \times \tilde{F}^{(n)}_{l,[2^{l-1} u]+1}(u)\Bigg]\ud u 
\end{eqnarray*}
together with Eqs.~(\ref{eq:24}) and (\ref{eq:25}), it is not difficult to establish the following proposition.
\begin{Pr1}
\label{Pr:1}
Assume that the quadrature rule specified by the the points $0 = u_0 < u_1 \ldots < u_q = 1$ and the nonnegative weights $w_0, \ldots, w_q$  integrates exactly all the expressions of the form 
\begin{equation}
\label{eq:33}
\int_0^1 V\left[c_1 + c_2 u + c_3 C_0(u) + c_4 L_0(u) + c_5 R_0(u)\right]\ud u
\end{equation}
for a given class of continuous functions $V(x)$. Then the quadrature rule specified by the $q2^k +1$ points 
\[u'_p =
 \left(u_{p-q[p/q]} + [p/q]\right)2^{-k}, \quad  \text{for}\ 0 \leq p \leq q2^k\]
and the corresponding weights 
\[w'_p =\left\{\begin{array}{l l}
w_0 2^{-k}, &\text{for} \ p = 0 \\ w_q 2^{-k}, &\text{for} \ p = q2^k, \\
(w_0 + w_q) 2^{-k}, &\text{for} \ p = q j,\; 0 < j < 2^k, \\ w_{p-q[p/q]} 2^{-k}, &  \text{otherwise} \end{array}\right.\]
integrates exactly all the expressions of the form
\[\int_0^1 V\left[x_r(u)+\sigma \sum_{l=1}^{k+2} a_{l,[2^{l-1} u]+1} \;\tilde{F}^{(n)}_{l,[2^{l-1} u]+1}(u)\right]\ud u\]
for the same class of potentials.
\end{Pr1}
While the straightforward proof of the proposition is left for the reader, we mention that an alternative demonstration can be achieved by means of Eqs.~(\ref{eq:30}) and (\ref{eq:31}). Again, $[p/q]$ denotes the integer part of $p/q$.

Proposition~(\ref{Pr:1}) reduces the quadrature problem to the problem of integrating expressions of the type given by Eq.~(\ref{eq:33}), which are  simpler. A further reduction can be achieved by noticing that the function $C_0(u)$ is symmetrical under the transformation $u \to 1-u$, while the functions $R_0(u)$ and $L_0(u)$ transform one into each other. Moreover, any linear combination $c_0 + c_1 u$ remains a linear combination of the functions $1$ and $u$ under the same transformation. These arguments allow us to restrict our search to the interval $[0, 1/2]$ only, because the rest of the quadrature points on the whole interval $[0,1]$ can be obtained by symmetry. 
To integrate all the expressions of the form 
\[
\int_0^{1/2} V\left[c_1 + c_2 u + c_3 C_0(u) + c_4 L_0(u)\right]\ud u
\]
(the function $R_0(u)$ disappeared because it is zero on the interval $[0, 1/2]$) for all quadratic potentials, it is enough that the following system of equations be satisfied:
\begin{equation}
\label{eq:34}
\left\{\begin{array}{l l}
\sum_{i = 0}^{4}w_i = \int_0^{1/2}1 \ud u, \\
\sum_{i = 0}^{4}w_i u_i = \int_0^{1/2} u \ud u,\\
\sum_{i = 0}^{4}w_i u_i^2 = \int_0^{1/2} u^2 \ud u, \\
\sum_{i = 0}^{4}w_i C_0(u_i) = \int_0^{1/2} C_0(u) \ud u,\\
\sum_{i = 0}^{4}w_i C_0(u_i)^2 = \int_0^{1/2} C_0(u)^2 \ud u, \\ 
\sum_{i = 0}^{4}w_i u_i C_0(u_i) = \int_0^{1/2} u C_0(u) \ud u,\\
\sum_{i = 0}^{4}w_i L_0(u_i) = \int_0^{1/2} L_0(u) \ud u, \\
\sum_{i = 0}^{4}w_i u_i L_0(u_i) = \int_0^{1/2} u L_0(u) \ud u,\\ 
\sum_{i = 0}^{4}w_i L_0(u_i) C_0(u_i) = \int_0^{1/2} L_0(u)C_0(u) \ud u. 
\end{array}\right.
\end{equation}
Notice that the condition
\begin{equation}
\label{eq:35}
\sum_{i = 0}^{4}w_i L_0(u_i)^2 = \int_0^{1/2} L_0(u)^2 \ud u
\end{equation}
was not introduced in the above system of equations because it is not independent of the  others. More precisely, we have the relation
\[
C_0(u)^2 + L_0(u)^2 = u(1-u),
\]
which shows that the equality~(\ref{eq:35}) is automatically satisfied provided that the first three and the fifth equalities from the above system hold.

The system given by Eq.~(\ref{eq:34}) contains $9$ equations and this is why the minimal number of quadrature points for the interval $[0,1/2]$ is $5$. Consequently, the number of unknown variables is $10$ and so we can arbitrarily fix one of the variables $u_i$. We have chosen to set $u_0 = 0$. With this choice, the system of equations described by Eq.~(\ref{eq:34}) can be solved by means of the Levenberg-Marquardt algorithm as implemented in Mathcad.\cite{Mathcad}  The computed quadrature points and weights together with their extension to the whole interval $[0,1]$ are shown in Table~\ref{Tab:1}. Luckily enough, the weights are all positive and therefore the quadrature scheme is numerically robust. 

\begingroup
\squeezetable
\begin{table*}[!bthp]
\caption{
\label{Tab:1}
Quadrature points and weights for the RW-LCPI method.}
\begin{tabular}{|c |c |c |c |c |c |c |c |c |c |c |}
\hline
$i$ & 0 & 1 & 2 & 3 & 4 & 5 & 6 & 7 & 8 & 9 \\ 
\hline \hline
$u_i$&0.000000000&0.081468437&0.232499762&0.36369423&0.468289398& 0.531710602&0.63630577&0.767500238&0.918531563&1.000000000\\
\hline
$w_i$&0.014247669&0.149656564&0.131280659&0.130290452&0.074524656&               	 0.074524656&0.130290452&0.131280659&0.149656564&0.014247669\\
\hline 
\end{tabular}
\end{table*}
\endgroup

It is instructive to verify numerically the findings of the present subsection for a one-dimensional system made up of a particle of mass $m_0 = 1$ moving in the quadratic potential $V(x) = m_0\omega^2 x^2/2$ of frequency $\omega =1$. We use atomic units (therefore, $\hbar = 1$) and we set $\beta = 10$. As discussed in the preceding section, it is more convenient to study the convergence of the discrete path integral method based on the L\'evy-Ciesielski reweighted technique because this can be performed by numerical matrix multiplication and it coincides with the corresponding series representation for $n = 2^k -1$. 

By numerical matrix multiplication, we compute a sequence of partition functions $Z_{2m+1}^{\text{LC}}(\beta)$ and then we compute the ratios
\[
R_{2m+1}^{\text{LC}}(\beta) = Z_{2m+1}^{\text{LC}}(\beta) \big/ Z(\beta)
\]
as well as the quantities
\[
\alpha_m^\text{LC} = m^2 \ln\left[1+\frac{R_{2m-1}^{\text{LC}}(\beta)-R_{2m+1}^{\text{LC}}(\beta)} {R_{2m+1}^{\text{LC}}(\beta)- 1}\right].
\]
Here, $Z(\beta)$ denotes the exact partition function. 
Assuming that the asymptotic convergence of the odd sequence $R_{2m+1}^{\text{LC}}(\beta)$ can be described by the equation
\begin{eqnarray}
\label{eq:36}\nonumber
R_{2m+1}^{\text{LC}}(\beta)= 1 + \frac{c_{\text{LC}}}{(2m+1)^\alpha} + \frac{c'_{\text{LC}}}{(2m+1)^{\alpha +1}}\\ +o\left(\frac{1}{(2m+1)^{\alpha+1}}\right),
\end{eqnarray}
it is not difficult to show that the slope of $\alpha_{m}^\text{LC}$ as a function of $m$ converges to the convergence exponent $\alpha$ as fast as $o(1/m)$.\cite{Pre02} Moreover, once the exponent $\alpha$ is determined, one may compute the quantities
\[
c_m^{\text{LC}} = (2m+1)^\alpha m \left[R_{2m+1}^{\text{LC}}(\beta) - 1\right]
\]
and show that $c_{m+1}^\text{LC}-c_m^\text{LC}$ converges to $c_\text{LC}$ as fast as $o(1/m)$. 
A similar discussion is true of the even sequence $R_{2m}^{\text{LC}}(\beta)$, though for the case of the Wiener-Fourier reweighted technique, the constant $c'_\text{LC}$ appearing in Eq.~(\ref{eq:36}) has usually different values for the odd and the even subsequences. To avoid the appearance of certain oscillations in our plots as well as to maintain a unified exposition of the methods employed, we study only the odd subsequence but mention that the even subsequence was also studied and produced the same results. 

The short-time high-temperature approximation utilized in the NMM technique has the expression
\begin{widetext} 
\begin{eqnarray}\nonumber
\label{eq:37}
\frac{\rho_0^{\text{RW}}(x,x';\beta)}{\rho_{fp}(x,x';\beta)}=\int_{\mathbb{R}} \ud \mu(a_1)\Bigg[ \int_{\mathbb{R}}\ud \mu(a_2)  \exp\bigg\{-\beta  \sum_{i = 0}^4 w_i V[x +(x'-x)u_i+a_1\sigma C_0(u_i)  +a_2 \sigma L_0(u_i)] \bigg\} \Bigg]\times \\ \Bigg[ \int_{\mathbb{R}}\ud \mu(a_2)  \exp\bigg\{-\beta \sum_{i = 5}^9 w_i V[x+(x'-x)u_i+a_1\sigma C_0(u_i)  +a_2 \sigma R_0(u_i)]\bigg\} \Bigg],  
\end{eqnarray}
\end{widetext}
where the one-dimensional integrals appearing in Eq.~(\ref{eq:32}) have been transformed into finite sums with the help of the quadrature points tabulated in Table~\ref{Tab:1}. As discussed in the preceding section, the Gaussian integrals are computed by Gauss-Hermite quadrature in $10$ points for each dimension. We have also computed a sequence ${\alpha'}^\text{LC}_m$ for which the $10$ quadrature points from Table~\ref{Tab:1} were replaced by the same number of Gauss-Legendre quadrature points.\cite{Pre92}

\begin{figure}[!tbp] 
   \includegraphics[angle=270,width=8.5cm,clip=t]{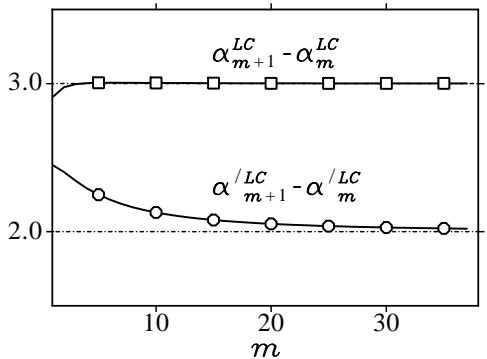} 
 \caption[sqr]
{\label{Fig:3}
The convergence order of the RW-LCPI method depends upon the nature of the quadrature scheme employed.
}
\end{figure} 

As shown in Fig.~\ref{Fig:3}, the sequence $\alpha_{m+1}^\text{LC} - \alpha_m^\text{LC}$ converges to $3$ as opposed to ${\alpha'}^\text{LC}_{m+1} - {\alpha'}^\text{LC}_m$, which converges to $2$. This demonstrates that both the number of quadrature points and the nature of the quadrature technique play an important role for the value of the convergence order. In fact, additional numerical examples not presented here show that doubling the number of quadrature points for the Gauss-Legendre method does not improve the convergence order. We have also tested the trapezoidal quadrature scheme and found again a convergence order of $2$. Therefore, the quadrature scheme is important and it cannot be chosen arbitrarily.  We therefore recommend the use  of the quadrature scheme developed in the present section for practical applications. In the following section, we will demonstrate by numerical examples that this scheme preserves the cubic convergence of the RW-LCPI method for all smooth enough potentials. 

\begin{figure}[!tbp] 
   \includegraphics[angle=270,width=8.5cm,clip=t]{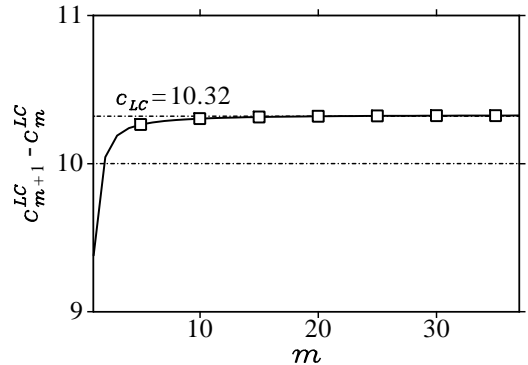} 
 \caption[sqr]
{\label{Fig:4}
The extrapolated value of the convergence constant for the RW-LCPI method is $c_\text{LC} = 10.32$.
}
\end{figure}

In Fig.~\ref{Fig:4}, we plotted the sequence $c_{m+1}^\text{LC}-c_m^\text{LC}$, the limit of which is the convergence constant $c_\text{LC} \approx 10.32$. This value of the convergence constant will be compared in the following subsection with the one for the Wiener-Fourier reweighted technique. As discussed in Section~II, the Wiener-Fourier reweighted technique is expected to have a smaller convergence constant.

\subsection{Minimalist quadrature rule for the RW-WFPI method}
Theoretically speaking, we may use a strategy similar to the one employed for the RW-LCPI method in order to develop a quadrature technique which integrates exactly all the quadratic potentials. However, for the RW-WFPI method, this approach is rather problematic because it requires solving large dimensional non-linear systems of equations of the type shown by Eq.~(\ref{eq:34}). Also, because the products of the functions utilized for path expansion present a strong linear dependence even after all functional relations are taken into account, the resulting system of equations is rather ill-conditioned. 

The alternative approach advocated in the present subsection is an empirical one. Let us remember that the functions utilized in the path expansion are $1$, $u$, $\sin(k\pi u)$ for $ 1\leq k \leq n$, and $r_n(u)\sin(k\pi u)$ for $n < k \leq 4n$. It is known that $\lim_{n \to \infty} r_n(u) = 1$.\cite{Pre03} It follows that these path expansion functions as well as their products are highly oscillatory functions which may be well approximated by the Legendre orthogonal polynomials on the interval $[0,1]$. This might be so because the latter ones are also highly oscillatory. This suggests that the Gauss-Legendre quadrature scheme (which we have found does not perform well for the RW-LCPI method) may be a good quadrature scheme for the RW-WFPI method.

Before moving on to numerical examples, let us define what we mean by ``good'' quadrature schemes. We require that the quadrature scheme preserve the asymptotic convergence of the RW-WFPI method. More precisely, both the order of convergence and the convergence constant should be preserved. Let us define 
\begin{equation}
\label{eq:38}
c_m^{\text{WF}}= (2m+1)^3 \frac{Z_{2m+1}^\text{WF}(\beta)- Z(\beta)}{Z(\beta)}.
\end{equation} 
Again, we study the convergence of the odd subsequence, but mention that the even subsequence was also studied and produced similar results. From Eq.~(\ref{eq:15}), one computes the following theoretical value for the convergence constant
\begin{eqnarray}
\label{eq:39}&& \nonumber 
c_\text{WF}^\text{th}= \lim_{m \to \infty} c_m^{\text{WF}}= -
 \frac{2\hbar^2\beta^3}{N_{0}\pi^4 m_0}\\ &&  \times \frac{1}{Z(\beta)} 
\int_\mathbb{R}\rho(x,x;\beta)V'(x)^2 \ud x \approx -0.684.
\end{eqnarray}

\begin{figure}[!tbp] 
   \includegraphics[angle=270,width=8.5cm,clip=t]{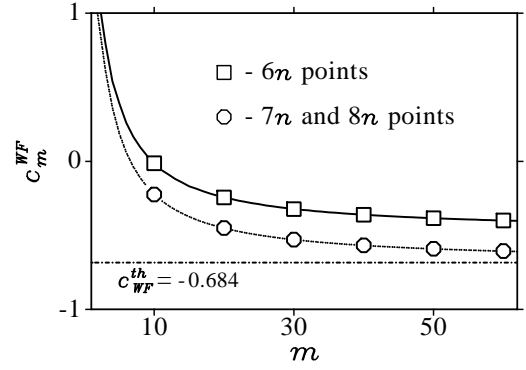} 
 \caption[sqr]
{\label{Fig:5}
The convergence constant for the RW-LCPI method depends upon the number of quadrature points in the Gauss-Legendre scheme. Using $7n = 7(2m+1)$ quadrature points or more is enough to recover the correct theoretical constant. 
}
\end{figure} 

The numerical evaluation of the sequence $c_m^{\text{WF}}$ by matrix diagonalization has been discussed in Appendix~C of Ref.~(\onlinecite{Pre03}). Employing the method developed there, we have computed numerical approximations to the terms $c_m^{\text{WF}}$ using $6n$, $7n$, and $8n$ Gauss-Legendre quadrature points, respectively (remember that $n = 2m+1$). The results plotted in Fig.~\ref{Fig:5} clearly demonstrate that the convergence order is  $3$ for all three cases. However, one notices that if the number of quadrature points is larger or equal to $7n$, the results are virtually indistinguishable and the computed values converge to one and the same limit. This limit is automatically the theoretical convergence constant.

\begin{figure}[!tbp] 
   \includegraphics[angle=270,width=8.5cm,clip=t]{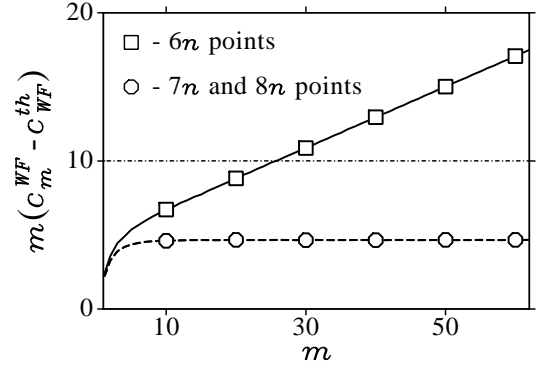} 
 \caption[sqr]
{\label{Fig:6}
This plot shows the qualitative difference for going from $6n$ to $7n$ or more quadrature points. In the latter case, the theoretical convergence constant of the RW-WFPI method is preserved. 
}
\end{figure} 

We can reinforce the findings of the preceding paragraph by studying the convergence of the sequence $m\left(c^\text{WF}_m - c^\text{th}_\text{WF}\right)$. If the sequence is convergent, then 
\[\lim_{m \to \infty} c^\text{WF}_m = c^\text{th}_\text{WF}. 
\]
On the other hand, if the asymptotic dependence of $m\left(c^\text{WF}_m - c^\text{th}_\text{WF}\right)$ with $m$ is linear with a non-zero slope, then
\[\lim_{m \to \infty} c^\text{WF}_m \neq c^\text{th}_\text{WF}. 
\]
The results plotted in Fig.~\ref{Fig:6} clearly demonstrate that the asymptotic convergence of the RW-WFPI method (meaning both the convergence order and the convergence constant) is preserved provided that the number of Gauss-Legendre quadrature points is $7n$ or larger.

We conclude this section by performing a comparison between the rates of convergence of the  RW-LCPI and RW-WFPI methods. For a given $n$, the former method employs $4n+3$ Gaussian parameters for path expansion, while the latter method employs $4n$. The two numbers can be thought of as equal for large $n$. However, the modulus of the convergence constant for the RW-WFPI method is $0.684$, almost $15$ times smaller than the convergence constant for the RW-LCPI method, which is $10.32$. As order of magnitude, the difference between the two constants  at low enough temperatures can be understood as follows. As shown by Eq.~(\ref{eq:39}), the dependence of the convergence constant with $\beta$ at low temperature is of the form
\[
c^\text{th}_\text{WF}(\beta) \propto \beta^3. 
\]
On the other hand, assuming that the convergence order for the RW-LCPI method is $3$ for all smooth enough potentials, the short-time high-temperature approximation given by Eq.~(\ref{eq:31}) must be of the form
\[
\rho_0^\text{LC}(x,x';\beta) = \rho(x,x';\beta)\left[1 + O(\beta^4)\right] 
\]
for small enough $\beta$. This is so because the convergence order of the discrete path integral method described by Eq.~(\ref{eq:30}) and the order of the short-time approximation given by Eq.~(\ref{eq:31}) are always related by the following Suzuki formula\cite{Suz85}
\[
\rho_n^\text{LC}(x,x';\beta) = \rho(x,x';\beta)\left[1 + O(\beta^4/n^3)\right]. 
\]
This formula shows that 
\[
c^\text{th}_\text{LC}(\beta) \propto \beta^4 
\]
and explains why the Wiener-Fourier reweighted technique performs better than its L\'evy-Ciesielski version at low temperature (large $\beta$). 

\section{More numerical examples}
In this section, we present numerical evidence supporting the idea that the convergence of the reweighted techniques described in the previous section is indeed cubic for smooth enough potentials even if only the proposed minimal quadrature techniques are employed. Strictly speaking, we will be able to verify our findings only for the RW-LCPI method by numerical matrix multiplication. For the RW-WFPI method, we shall compute the partition functions of the systems studied by Monte Carlo integration and show that they are better than the partition functions obtained by means of  the RW-LCPI method. Clearly,  this provides only indirect evidence for the asymptotic convergence of the RW-WFPI method. As discussed in Section~IV.B of Ref.~(\onlinecite{Pre02}), it is very difficult to compute the partition functions by Monte Carlo integration with accuracy good enough to  determine unequivocally the convergence order of a method.  

Besides the convergence order, we shall also verify that the quadrature technique developed in Section~IV.A preserves the convergence constant of the RW-LCPI method. Since we do not have an explicit expression for it, we determine the convergence constant numerically by using a large number of Gauss-Legendre quadrature points to compute the ``exact'' short-time approximation given by Eq.~(\ref{eq:32}). Thus, we compute the sequences $\alpha_m^{\text{LC}}$ and $c_m^{\text{LC}}$ defined in Section~III.A with the help of the $10$ quadrature points from Table~\ref{Tab:1}, but also the sequences $\alpha'^{\text{LC}}_m$ and $c'^{\text{LC}}_m$ with the help of $500$ Gauss-Legendre quadrature points. In the preceding section, we suggested that no matter how large the number of Gauss-Legendre quadrature points is, the convergence order of the resulting method is at most $O(1/n^2)$. However, for a fixed range of values of $n$, one can obtain essentially exact results for the quantities $\alpha'^{\text{LC}}_m$ and $c'^{\text{LC}}_m$ by employing a large number of Gauss-Legendre quadrature points. Numerical experimentation shows that the optimal number is $500$ for the range of $n$ employed in the present study ($n = 2m+1$, $m = 1, 2, \ldots, 60$). 

\subsection{Quartic potential}

The first example we study is  the quartic potential $V(x)=x^4/2$. We set $\hbar =1$, $m_0=1$, and $\beta = 10$. With the help of the numerical matrix multiplication technique described in Section~III.A, we compute the quantities
\[
R_{2m+1}^{\text{LC}}(\beta) = Z_{2m+1}^{\text{LC}}(\beta) \big/ Z(\beta)
\]
and
\[
\alpha_m^\text{LC} = m^2 \ln\left[1+\frac{R_{2m-1}^{\text{LC}}(\beta)-R_{2m+1}^{\text{LC}}(\beta)} {R_{2m+1}^{\text{LC}}(\beta)- 1}\right].
\]
The exact partition function $Z(\beta)$ has been evaluated by variational methods and has the value
\[ Z(\beta) \approx 4.982570651\cdot 10^{-3}.\]
To compute the partition functions $Z_{2m+1}^{\text{LC}}(\beta)$ by numerical matrix multiplication, we have restricted the problem to the interval $[-4, 4]$. A division of this interval in $M = 1024$ slices is sufficient for  $Z_{2^{14}-1}(\beta)$ to reproduce the first $9$ digits of the exact partition function. 

As previously discussed,  the slope of $\alpha_{m}^\text{LC}$ as a function of $m$ converges to the convergence exponent $\alpha$. Once the exponent $\alpha$ is determined, one may obtain the convergence constant by analyzing the slope of
\[
c_m^{\text{LC}} = (2m+1)^\alpha m \left[R_{2m+1}^{\text{LC}}(\beta) - 1\right].
\]
A similar study is performed for the sequences $\alpha'^{\text{LC}}_m$ and $c'^{\text{LC}}_m$, the slopes of which converge to the exact convergence exponent and convergence constant, respectively. 

\begin{figure}[!tbp] 
   \includegraphics[angle=270,width=8.5cm,clip=t]{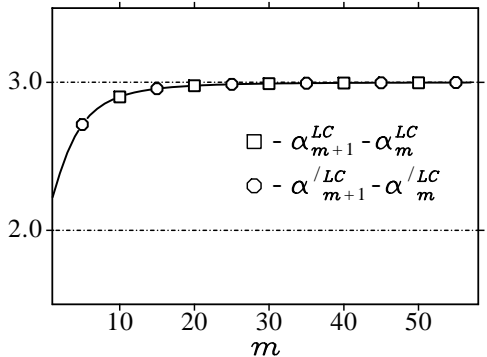} 
 \caption[sqr]
{\label{Fig:7}
This plot actually contains two superimposed graphs showing that the slopes of $\alpha'^{\text{LC}}_m$ and $\alpha^{\text{LC}}_m$ converge to $3$.
}
\end{figure}

\begin{figure}[!tbp] 
   \includegraphics[angle=270,width=8.5cm,clip=t]{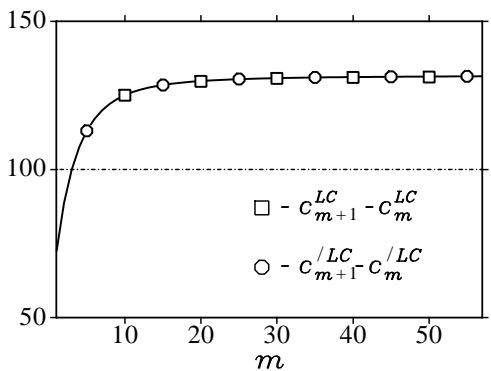} 
 \caption[sqr]
{\label{Fig:8}
This plot contains two superimposed graphs demonstrating that the slopes of $c'^{\text{LC}}_m$ and $c^{\text{LC}}_m$ are virtually indistinguishable. Therefore, the slopes converge to the same limit as $m \to \infty$.
}
\end{figure} 

As demonstrated by Figs.~\ref{Fig:7} and \ref{Fig:8}, the RW-LCPI method based on the quadrature technique developed in Section~IV.A and the exact RW-LCPI method have virtually identical convergence properties. Both methods have cubic convergence and have equal convergence constants. This is strong evidence supporting the hypothesis that if a quadrature technique preserves the asymptotic convergence of a reweighted method for the harmonic oscillator, then it preserves the asymptotic convergence for all smooth enough potentials. 

Even if we cannot perform a similar analysis for the RW-WFPI method, it is instructive to compare the values of the partition function with those computed by the trapezoidal Trotter and RW-LCPI methods. In Table~\ref{Tab:2}, we have listed the partition functions $Z^\text{TT}_{4n-1}(\beta)$, $Z^\text{LC}_{n-1}(\beta)$, and $Z^\text{WF}_{n}(\beta)$ for $n = 2^k, \; k=1,\ldots, 6$. By   computing $Z^\text{TT}_{4n-1}(\beta)$, $Z^\text{LC}_{n-1}(\beta)$, and $Z^\text{WF}_{n}(\beta)$, we ensure fair comparison as to the number of  variables used to parameterize the paths. The first two methods utilize a number of $4n-1$ path variables, while the last one utilizes $4n$ variables. One notices that even if the L\'evy-Ciesielski reweighted method has cubic convergence, the improvement over the trapezoidal Trotter is not that great for small numbers of path variables. This is due to the fact that the convergence constant of the RW-LCPI method is large, being proportional to $\beta^4$. The Wiener-Fourier reweighted method, which has a convergence constant proportional to $\beta^3$, produces much better values for a similar number of path variables. The RW-WFPI partition functions decrease up to a point ($n = 32$) and then increase back to the exact partition function. This is consistent with the fact that the convergence constant of the RW-WFPI method is negative.

\begingroup
\begin{table}[!bthp]
\caption{
\label{Tab:2}
Partition functions for the quartic potential. The partition functions for the RW-WFPI method were computed by Monte Carlo integration in 1 billion points. We employed $7n$ Gauss-Legendre quadrature points to compute the path averages.}
\begin{tabular}{|c |c |c |c |}
\hline
$n$ & $Z^\text{TT}_{4n-1}(\beta)$ &  $Z^\text{LC}_{n-1}(\beta)$ & $Z^\text{WF}_{n}(\beta)$ \\ 
\hline \hline
$2$ &$10.235\cdot 10^{-3}$&$10.688\cdot 10^{-3}$&$(7.899 \pm 0.003)\cdot 10^{-3}$\\
\hline
$4$ &$6.489\cdot 10^{-3}$&$6.538\cdot 10^{-3}$&$(5.518\pm 0.003)\cdot 10^{-3}$\\
\hline 
$8$ &$5.393\cdot 10^{-3}$&$5.364\cdot 10^{-3}$&$(5.047 \pm 0.002)\cdot 10^{-3}$ \\
\hline 
$16$&$5.089\cdot 10^{-3}$&$5.062\cdot 10^{-3}$&$(4.983 \pm 0.002)\cdot 10^{-3}$\\
\hline 
$32$&$5.009\cdot 10^{-3}$&$4.996\cdot 10^{-3}$&$(4.979  \pm 0.002)\cdot 10^{-3}$ \\
\hline 
$64$&$4.989\cdot 10^{-3}$&$4.985\cdot 10^{-3}$&$(4.982  \pm 0.002)\cdot 10^{-3}$ \\
\hline 
$\infty$&
\multicolumn{3}{|c|}{$4.983\cdot 10^{-3}$}\\
\hline
\end{tabular}
\end{table}
\endgroup

\subsection{ He cage problem}
The physical system\cite{Fre86} we consider in this subsection consists of a particle trapped on a line between two atoms separated by a distance $L$. The particle is assumed to interact with the fixed atoms through pairwise Lennard-Jones potentials. The resulting cage is described by the potential
\[
V(x)= 4\epsilon \left[\left(\frac{\sigma}{x}\right)^{12} - \left(\frac{\sigma}{x}\right)^{6}+ \left(\frac{\sigma}{x-L}\right)^{12} - \left(\frac{\sigma}{x-L}\right)^{6}\right],
\]
if $0 < x < L$ and $V(x) = +\infty$, otherwise. The parameters of the system are chosen to be those for the He atom. We set $m_0 = 4\; \text{amu}$, $\epsilon / k_{B} = 10.22 \; \text{K}$, $\sigma = 2.556 \stackrel{\circ}{\text{A}}$, and $L= 7.153 \stackrel{\circ}{\text{A}}$. At $T = 5.11 \;\text{K}$, which is the temperature utilized in the present computations, the system is practically in its ground state.  Numerical experimentation shows that in order to compute the partition functions $Z_{2m+1}^{\text{LC}}(\beta)$ by numerical matrix multiplication, it is safe to restrict the problem to the interval $[0.2L, 0.8L]$ and consider a division of this interval in $M = 1024$ equal slices.  By employing large values of $m$, the exact partition function was found to be \[Z(\beta) \approx Z_{2^{15}-1}^{\text{LC}}(\beta) = 1.668294002\ldots\]  

\begin{figure}[!tbp] 
   \includegraphics[angle=270,width=8.5cm,clip=t]{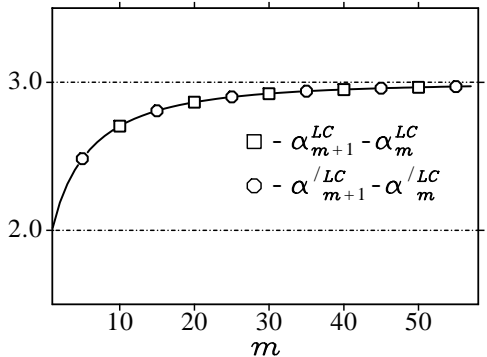} 
 \caption[sqr]
{\label{Fig:9}
This plot actually contains two superimposed graphs showing that the slopes of $\alpha'^{\text{LC}}_m$ and $\alpha^{\text{LC}}_m$ converge to $3$. This demonstrates that the convergence order of the RW-LCPI method for the He cage problem is cubic.
}
\end{figure}

\begin{figure}[!tbp] 
   \includegraphics[angle=270,width=8.5cm,clip=t]{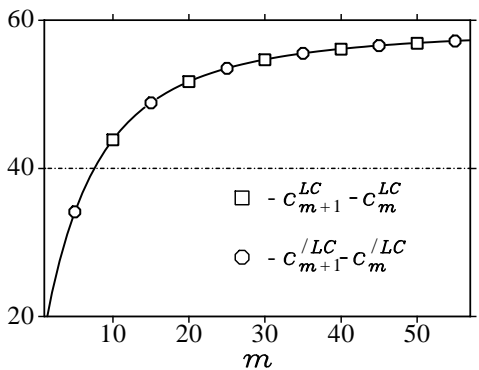} 
 \caption[sqr]
{\label{Fig:10}
This plot contains two superimposed graphs demonstrating that the slopes of $c'^{\text{LC}}_m$ and $c^{\text{LC}}_m$ for the He cage problem are virtually indistinguishable. Therefore, the slopes converge to the same limit as $m \to \infty$, which demonstrates that nothing is lost by employing the quadrature technique developed in Section~IV.A.
}
\end{figure}

Figs.~\ref{Fig:9} and \ref{Fig:10} demonstrate that the convergence of the RW-LCPI method for the He cage problem is cubic. Moreover, there is virtually no difference between the values computed using the quadrature technique of Section~IV.A and those computed using $500$ Gauss-Legendre quadrature points. Similar to the case of the quartic potential, the partition functions $Z_{n-1}^{\text{LC}}(\beta)$ and $Z_{4n-1}^{\text{TT}}(\beta)$ are very close one to each other for small $n$ even if the asymptotic convergence of the methods is different. On the other hand, as seen from Table~\ref{Tab:3}, the partition functions computed with the help of the Wiener-Fourier reweighted technique are significantly better. The asymptotic convergence of $Z_{n}^{\text{RW}}(\beta)$ is eventually from below, in agreement with the theoretical prediction for the exact Wiener-Fourier reweighted technique.  

\begingroup
\begin{table}[!bthp]
\caption{
\label{Tab:3}
Partition functions for the He cage problem. The partition functions for the RW-WFPI method were computed by Monte Carlo integration in 1 billion points and using $7n$ Gauss-Legendre quadrature points to compute the path averages.}
\begin{tabular}{|c |c |c |c |}
\hline
$n$ & $Z^\text{TT}_{4n-1}(\beta)$ &  $Z^\text{LC}_{n-1}(\beta)$ & $Z^\text{WF}_{n}(\beta)$ \\ 
\hline \hline
$2$ &$1.9161$&$1.8917$&$1.8084  \pm   0.0003 $\\
\hline
$4$ &$1.7597$&$1.7545$&$1.7050   \pm   0.0003 $\\
\hline 
$8$ &$1.6975$&$1.6958$&$1.6739  \pm   0.0003 $ \\
\hline 
$16$&$1.6766$&$1.6753$&$1.6686  \pm  0.0003 $\\
\hline 
$32$&$1.6705$&$1.6698$&$1.6681  \pm  0.0003$ \\
\hline 
$64$&$1.6688$&$1.6685$&$1.6683  \pm  0.0003$\\
\hline
$\infty$&
\multicolumn{3}{|c|}{$1.6683$}\\
\hline
\end{tabular}
\end{table}
\endgroup

The He cage problem is interesting because the Lennard-Jones potential lies outside the class of potentials for which the reweighted methods were proven to have cubic convergence.\cite{Pre03} The example studied suggests that  if $\exp[-\beta V(x)]$ has Sobolev derivatives of up to the second order for all $\beta > 0$ and if  the quantities 
\[ 
\int_{\mathbb{R}} \rho(x,x;\beta) V'(x)^2 \ud x
\]
and
\[ 
\int_{\mathbb{R}} \rho(x,x;\beta) V''(x)^2 \ud x
\]
are finite, then the reweighted methods have cubic convergence. (Remember that the expectation values of the squares of the first and second order derivatives control the convergence constants of the reweighted techniques.) In the case of the Lennard-Jones potential, $\exp[-\beta V(x)]$ has continuous derivatives of any order and the density matrix has an exponential decay near singularities (see also the Appendix). Consequently,  the expectation values of the first and second derivatives are finite. We believe the cubic convergence of the reweighted methods extends to all potentials satisfying these criteria.

\section{Conclusions}

In this article, we have discussed the numerical implementation of two special reweighted techniques based upon the L\'evy-Ciesielski and the Wiener-Fourier series, respectively. For both methods, we have developed  minimalist quadrature techniques capable of preserving the asymptotic convergence of the respective methods, while employing a number of quadrature points proportional to the number of path variables. Moreover, we have demonstrated by numerical examples that the asymptotic convergence of the methods is cubic for smooth enough potentials. The numerical evidence is consistent with the mathematical analysis previously performed  by one of us.\cite{Pre03} 

The findings of the present paper allow us to conclude that the reweighted techniques have an asymptotic convergence better than that of the trapezoidal Trotter DPI method or other related direct methods. More precisely, the reweighted L\'evy-Ciesielski method is the most efficient method if the calculation of the paths is the dominant computational factor (theoretically, this will always be the case for a large enough number of path variables $n$). The RW-LCPI method has cubic convergence and requires a number of operations proportional to $n\log_2(n)$ for the computation of the the paths. The Wiener-Fourier reweighted method is the most efficient direct method as to the minimization of the number of calls to the potential. This is so because the  convergence constant of the RW-WFPI method is smaller than that of the RW-LCPI method, especially at low temperature. Even if the number of operations necessary to compute the paths is proportional to  $n^2$,  the RW-WFPI technique will most likely be the most efficient method for practical applications. Calls to the potential routine are usually expensive, whereas the calculation of the paths takes full advantage of the vector floating point units available in modern processors. 

\begin{acknowledgments}
 The authors acknowledge support from the National Science Foundation
through awards CHE-0095053 and CHE-0131114. 
\end{acknowledgments}

\appendix
\section{Potentials with strong positive singularities}

In this section, we shall discuss the class of potentials that are bounded from below and have  strong positive singularities at origin of the form $r^{-q}$ with $q > 0$. This class of potentials contains the Lennard-Jones potential. We shall first discuss the case of three dimensional potentials and then go back to the one-dimensional Lennard-Jones potential discussed in Section~V.B. We focus mainly on the three-dimensional examples because these are of practical importance. As such, we consider a particle of mass $m_0 = 1$ moving in a spherical potential of the form
\[
V(r) = U(r) + r^2/2, 
\]
where $U(r)$ is a bounded from below potential that behaves at origin as  $r^{-q}$ and is otherwise an infinitely differentiable function decaying to zero at infinity. The term $r^2$ is added to confine the system about the origin so that the resulting Hamiltonian has discrete spectrum. 

The question we ask is about the decay of the density matrix at the singularity $r = 0$. For the sake of having an example, we set $U(r) = r^{-q}$. Clearly, the classical diagonal density matrix $\exp[-\beta V(r)]$ has an exponential decay to zero at the singularity any time $q > 0$.
As the quantum diffusion takes place (we interpret $\beta$ as a time parameter), the system starts to explore the singularities of the potential and the diagonal density matrix tends to spread out. The strong decay to zero of the classical density matrix near singularities is attenuated and, as we shall see, if the singularities are not very strong, the quantum density matrix may even fail to decay to zero. On physical grounds, we expect the effect to get stronger in the limit $\beta \to \infty$, when the system is also brought into its ground state eigenfunction.  Therefore, it is not farfetched to assume that the decay of the density matrix at the singularity is controlled by the decay of the ground state eigenfunction. For this reason, we shall focus our attention on the behavior of the ground state eigenfunction near singularities. We remind the reader that the ground state eigenfunction satisfies the Schr\"odinger equation
\[
-\frac{1}{2} \frac{1}{r^2}\frac{\ud}{\ud r}\left[r^2 \frac{\ud}{\ud r} \Psi_0(r) \right] + V(r)\Psi(r)= E_0 \Psi(r)
\]
and that it is the only positive eigenfunction (one uses this property to easily check if an eigenfunction is the ground state eigenfunction). 

It has been discussed in Section~II.B of Ref.~(\onlinecite{Pre03a}) that if $q < 2$, then the potential $V(r)$ is a Kato-class potential. In this case, it is known that the density matrix \emph{does not} decay to zero at the singularity. A quick way to verify this is to notice that if $q < 2$, then the potential $V(r)$ has finite Gaussian transform. In this conditions, Theorem~4 of Ref.~(\onlinecite{Pre03b}) says that 
\[
\mathbb{E}\int_0^1 V[x_r(u)+ \sigma B_u^0 ] < \infty, \quad \forall \; (x, x') \in \mathbb{R}^3 \times \mathbb{R}^3
\] 
and for all $\beta > 0$. The last inequality together with Jensen's inequality implies
\begin{eqnarray*}
\frac{\rho(x,x';\beta)}{\rho_{fp}(x,x';\beta)} = \mathbb{E} \exp\left\{-\beta \int_0^1 V[x_r(u)+ \sigma B_u^0 ]\right\} \\ \geq \exp\left\{-\beta \mathbb{E}\int_0^1 V[x_r(u)+ \sigma B_u^0 ]\right\} > 0,
\end{eqnarray*}
which proves our assertion since the density matrix of a free particle is strictly positive.

On the other hand, if $q = 2$, the density matrix has polynomial decay to zero at $r = 0$, as shown by the decay  of the (unnormalized) ground state eigenfunction
\[\Psi_0(r)= r^{\frac{1}{2}\left(\sqrt{1+8\gamma}-1\right)}e^{-r^2/2} \]
of the potential
\[
V(r) = \frac{1}{2} \left(\frac{2\gamma}{r^2}+ r^2\right).
\]
The corresponding eigenvalue is $1 + \sqrt{1+8\gamma} / 2$. We notice that the rate of decay increases as  $\gamma$ becomes larger because the ``intensity'' of the positive singularity $\gamma r^{-2}$ also increases.

If $q > 2$, then the decay at the singularity is exponential. This is demonstrated by the decay at origin of the ground state eigenfunction
\[
\Psi_0(r) = \frac{1}{r} \exp\left[ -\left(\frac{1}{r^\gamma} + \frac{r^{2}}{2}\right)\right]
\]
of the potential
\[
V(r) =\frac{1}{2}\left\{ \frac{\gamma^2}{r^{2(\gamma + 1)}} -\frac{\gamma (\gamma + 1)}{r^{2+\gamma}} - \frac{2}{r^\gamma} + r^2\right\}, 
\]
where $q = 2(\gamma + 1)$. The ground state eigenvalue for the given example is $0.5$.

As the last example shows, the density matrix for a Lennard-Jones potential, potential that has a positive singularity of the form $r^{-12}$ at origin, is decaying exponentially fast at $r=0$. Because of this strong decay, the expectation values of the squares of the first and second derivatives of the potential against the diagonal density matrix are finite. As discussed in Section~V.B, in these conditions we expect that the convergence rates of the reweighted techniques are cubic. A similar discussion holds for a one-dimensional system. For instance, the ground state eigenfunction of the potential 
\[
V(x) = \left\{ \begin{array}{c l}\frac{1}{2}\left\{ \frac{\gamma^2}{x^{2(\gamma + 1)}} -\frac{\gamma (\gamma + 1)}{x^{2+\gamma}} - \frac{2}{x^\gamma} + x^2\right\}, & x > 0, \\ +\infty, & x \leq 0\end{array} \right.
\]
is 
\[
\Psi_0(x) = \left\{ \begin{array}{c c}\exp\left[ -\left(\frac{1}{x^\gamma} + \frac{x^{2}}{2}\right)\right], & x > 0, \\ 0, & x \leq 0\end{array} \right. .
\]
Again, $\Psi_0(x)$ has exponential decay at $x = 0$ whenever $\gamma > 0$. (To recover the singularity of the Lennard-Jones potential, set $\gamma = 5$).


\begin{thebibliography}{99}
\bibitem{Footnote}{The reader is remainded that a convergent sequence $x_n \to x$ has $o(1/n^s)$ asymptotic convergence if $\lim_{n \to \infty} n^s (x_n-x)=0$ and it has  $O(1/n^s)$ asymptotic convergence if there is a constant $M>0$ such that $ n^s |x_n-x|\leq M$ for all $n \geq 1$. In the latter case, if it exists and is not zero, the limit $\lim_{n \to \infty} n^s (x_n-x)$ is called the convergence constant of the sequence $x_n$. }
\end{thebibliography}
\end{document}